\documentclass[10pt]{amsart}
\usepackage{comment,balance,pdfpages}
\usepackage{soul}
\usepackage{colortbl}
\usepackage{dsfont}
\usepackage{amsthm}
\usepackage{stmaryrd}
\usepackage{amsmath,amssymb,amsfonts,bbm,multicol,caption}
\usepackage{graphicx}
\usepackage{tcolorbox}
\usepackage{xcolor,braket}
\usepackage{multirow,flushend}
\usepackage{array} 
\usepackage{longtable}
\usepackage{url}
\usepackage{tabularx} 
\usepackage{algorithm,algorithmic}
\usepackage{float}
\usepackage{tabularx}
\usepackage{booktabs}

\usepackage{tikz}
\usetikzlibrary{positioning, shapes, arrows}

\renewcommand{\P}{\mathbb{P}}

\usepackage{mathrsfs}
\usepackage[scr=rsfs,cal=boondox]{mathalfa}

\PassOptionsToPackage{hyphens}{url}
\usepackage{xurl}
\usepackage{acronym}
\usepackage[acronym]{glossaries}
\usepackage{glossary-mcols}

\begin{document}
	\title{Entanglement distribution protocols under imperfect fidelity and quantum memory conditions}
	\author{Claire Mesny}
   \address[Claire Mesny]{Orange Innovation}
\curraddr{2, Avenue Pierre Marzin, F-22300 Lannion}
\email{claire.mesny@orange.com} 
  \thanks{Presented at the Second Workshop on Workshop on Quantum Networked Applications and Protocols (QuNAP 2026), organized in conjunction with IEEE International Conference on Computer Communications, May 18, 2026}

\author{Fabrice Guillemin}
\address[Fabrice Guillemin]{Orange Innovation}
\curraddr{2, Avenue Pierre Marzin, F-22300 Lannion}
\email{fabrice.guillemin@orange.com}

        \author{Claire Goursaud}
   \address[Claire Goursaud]{INSA Lyon}
\curraddr{Institut National des Sciences Appliquées de Lyon, F-69000 Lyon}
\email{claire.goursaud@insa-lyon.fr}     

\keywords{Quantum networks; Entanglement distribution; Routing}

\date{\today}
	
\begin{abstract}
		
		The rapid development of quantum computers and sensors urges for the development of a quantum Internet capable of transmitting quantum bits over long distances. Photons used for quantum data transfer are  fragile over time and sensitive to their environment, so that they cannot be directly used over long distances. To remedy this problem, long distance paths are segmented into shorter links and entangled pairs of photons are distributed over these links and swapped to create end-to-end entangled pairs over long distances, eventually used for teleportation. In this paper, we develop an existing protocol taking account of fidelity and imperfect memories. We shorten the execution time and thus increase its link success probability creating the so-called Locally Heralded Distribution (LHD). It turns out that the proposed protocol outperforms some previous protocols. We benchmark through simulation the performances of protocols considered in this paper by using a blind entanglement protocol as a baseline. 
	\end{abstract}
\maketitle

	\section{Introduction}
	
	The emergence of a quantum Internet in the near future is today very likely thanks to intensive research and recent breakthroughs in quantum technology and the development of applications (for instance quantum sensor  \cite{Komar_2014}). Different architectures are under consideration and  follow  either a  stack models \cite{Ahmed:2025,Wehner18,Beauchamp25} or horizontal models \cite{Pirker2025}; various test-beds are being deployed to evaluate these architectures. 
	
	The basic principles of a quantum Internet relies on the use of entanglement, swapping and teleportation. As a matter as fact, a quantum bit (qubit), most of the time a photon, cannot be directly transmitted over long distances (say, more than 15 km) across an optical fiber because it can simply be lost or the carried quantum information is altered because of the interaction of the photon with the medium. To overcome the problem of long distances, the usual method consists of splitting a long distance path into smaller segments (say, 15 km) connected by nodes and to distribute entangled pairs over these segments; half pairs received at a node are swapped and the resulting Bell State Measurement (BSM) when successful is transmitted to the path endpoints. Under the assumption that along a path, the distribution  of entangled pairs over the successive links and the BSMs at nodes are successful, it is possible to create an end-to-end entangled pair, which is used to teleport a qubit from the source to the destination. 
	
	To implement the above scheme, it is necessary to store photons at intermediate nodes. Quantum memories seem today feasible \cite{Wehner18} and stored photons are used to create end-to-end entanglement. The choice of swapping half pairs distributed over contiguous links is under the  control of the routing protocol, which establishes routes through the network to connect an ingress and egress node. Because of the fragility of photons and limited performance of swapping operations, routing is preferably based on shortest path. The originality of the routing algorithm is that it should account of different factors: swapping probability, entangled pair distribution probability, fidelity of the end to end entangled pair, etc. 

    Before designing a routing protocol, it is utmost important to understand the performance of the entangled pair distribution protocol. In this paper we use one of the protocols introduced in \cite{Abane25}. We include imperfect memories and fidelity as a function of memory and fiber-induced decoherence, which is rarely covered at the same time in other works. We observe how much time its execution needs and how it impacts the fidelity. On the basis  of this protocol, called Globally Heralded Distribution (GHD), inspired from the Unheralded Distribution in \cite{Abane25}, we design a faster protocol which we call Locally Heralded Distribution (LHD).
	
	The organization of this paper is as follows: In Section~\ref{part:related}, we give a short overview of the literature regarding quantum routing and entanglement distribution and we introduce the basic concepts for entanglement distribution in Section~\ref{part:basics}. The protocols we use as baseline, referred to as  Blind Swapping, and the Globally Heralded Distribution are described in Section \ref{part:BS_GHD}, with emphasis on the link probability with imperfect memories. We also establish an analytical function for the link fidelity. Finally, the LHD protocol  is described in Section~\ref{part:LHD}. The performance of the protocols in terms of link fidelity and link success probability form all three protocols is analyzed in Section~\ref{performance}. Concluding remarks are presented in Section~\ref{conclusion}. 
	
	\section{Quantum routing and distribution protocols}
	\label{part:related}

	Directly sending a qubit through a medium is highly inefficient \cite{Pirandola_2017}. A routing algorithm in a network of quantum nodes aims to find a path between an ingress and egress node along which to establish an end-to-end entanglement. Each possible path has a link fidelity and a entangled pair distribution success probability (link probability) that are used, among others, to build the routing algorithm.

	These parameters are determined from the distribution protocol used on a given path between the source and destination nodes. Existing protocols aim to increase link probability, as in the works from \cite{Shi19routing}, \cite{Abane25}, and \cite{Wang24_SES_noF_nor_mem} that compares two different kinds of protocols. Others focus on fidelity such as \cite{Duer99} that focuses on nested purification and \cite{Zang_2023_F_decoh_mem} that simulates a hop-by-hop protocol or \cite{Chakraborty_2020} that centers around LP formulation. The work in \cite{Hughes:2025_asynchro_SES_<_PES} studies quantum memories induced dephasing in fidelity, but the above cited papers do not explicitly express link probability as a function of time spent in memory and fiber, neither do they consider both fiber and memory-induced decoherence in the expression for the fidelity.
	
	
	Quantum routing protocols build on link parameters to find the best path in different ways. Reviews have been published in the past few years with different classifications for quantum routers \cite{Kar2023,Dupuy23,Abane25}. Following the classification in \cite{Abane25}, some protocols are proactive, so that the actual entanglement distribution happens once the path has been decided \cite{Dupuy23,chakraborty2019}. 
	
	Other protocols use reactive routing in which the distribution happens at the same time as the routing is decided \cite{chakraborty2019}. Mixes of these classes have been studied in \cite{Gyongyosi_2019,Gyongyosi_2018,Gyongyosi_2017} and a virtual network can be created using left-over links from prior distribution protocols \cite{Shi2024_QCAST}.
	
	An established link has to follow a certain Quality of Service (QoS). Existing protocols aim to satisfy that QoS using different means, including fairness \cite{Kumar_Vinay_2025} and fidelity. Such protocols that focus on fidelity use the estimated link fidelity of a path in their cost function such as \cite{Kumar_Vinay_2025,li22fidel,ZhaoY2022,Chakraborty_2020}. Some even go further and consider the cost and possibility of using purification in the routing layer \cite{Patil2024,Duer99,Coopmans2024,li22fidel,Kumar_Pankaj_2025}.
	
	This paper focuses on entanglement distribution protocols, taking into consideration fidelity and time dependency in imperfect memories. We explicitly give link probability for the chosen protocols considering fiber and quantum memories parameters and we derive a complete end-to-end fidelity formula from not only swapping imperfections but also fiber and quantum memories interactions. 
		
	\section{Some physical realities of entanglement distribution}
	\label{part:basics}
	\subsection{Optical fiber transmission}
	
	The attenuation at telecom wavelength of $1550nm$ in typical telecommunication fibers is at best $\alpha^* = 0.14dB/km$. A specialized architecture of fiber for quantum signal, Hollow Core fiber (HCF), has been under study for years and achieves and attenuation as low as $0.09dB/km$ at telecom wavelength. When transmitting a photon through a fiber, the loss probability of the photon follows an exponential law with parameter depending on the distance and the attenuation parameter $\alpha^*$ so  that  for a length $L$ of the fiber, the loss probability of a photon is
	\begin{equation}
		1 - p_{\alpha} = 1 - e^{-L\alpha},\label{eq:p_loss}
	\end{equation} 
	where $p_{\alpha}$ is the success probability and $\alpha = \alpha^*/10$. The exponential decrease of $p_\alpha$ makes it impossible to directly send a photon though an optical fiber over long distances. Instead, we proceed with short distances and swap operations at intermediary nodes.
	
	Entangled pairs (ebits) are inevitably subjected to decoherence. Let the ideal ebit be in the Bell state $\ket{\Phi^+}$. Since it is affected by depolarizing noise, it may be decomposed into a Werner state $\rho$ with fidelity $F = \bra{\Phi^+}\rho\ket{\Phi^+}$ given by
	\begin{equation}
		\rho = \frac{4F-1}{3}\ket{\Phi^+}\bra{\Phi^+} + \frac{1-F}{3}\mathds{1}.\label{eq:werner_state}
	\end{equation} 
	
	For an initial fidelity $F$ and a coherence time $\tau_{coh}$, we assume that after  a time $\delta t$ the fidelity of the pair becomes \cite{Inesta_2023}
	\begin{equation}
		f_{\tau_{coh}}(\delta t, F) = \frac{1}{4} + (F - \frac{1}{4})e^{-\delta t/\tau_{coh}}.\label{eq:F_decoh}
	\end{equation} 
	
	The fidelity requirements depend on the application for the link, whether it is for QKD (Quantum Key Distribution) or distributed computing. QKD requires a lower threshold, around $90\%$, while distributed computing is more stringent.
	
	\subsection{Imperfections in quantum memories}
	
	Quantum memories are  starting to be commercialized \cite{Welinq_QM} and still present many challenges. For simplification, we suppose in this work that the write-in (conversion photon to memory qubit) and read-out (memory to photon) operations of the memories are deterministic; even if in practice, they are not and significantly lower the performance of memories.
	
	A qubit stored in memory, also known as a memory qubit, is also subjected to many  sources of error and follows a loss probability  similar to that of a  qubit crossing a fiber, namely an exponential law of the form
	\begin{equation}
		1 - p_{mem} = 1 - e^{-\delta t/\tau}\label{eq:p_mem}
	\end{equation} where $\delta t$ is the time spent in memory and $\tau$ is the qubit life-time in memory. The qubit memory life-time can go up to $\tau = 200 \ ms$. The ebit fidelity drops to $f_{\tau}(\delta t,F)$ given by Equation~\eqref{eq:F_decoh}.
	
\begin{figure}[htpb]
		\centering
		\includegraphics[width=.9\linewidth]{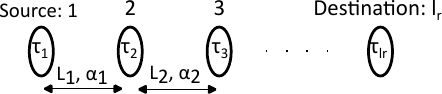}
		\caption{Path scheme with distance parameters $L_i$, attenuation parameter $\alpha_i$, and memory life-times $\tau_i$.}\label{fig:scheme_intro}
	\end{figure}

	We represent a path chosen for link establishment as illustrated in Figure~\ref{fig:scheme_intro}. The first node is the source and the last node, labeled $l_r$, is the destination. There are hence $l_r-1$ links with attenuation parameters $\alpha_i$ and lengths $L_i$ for each $i\in\llbracket 1,l_r-1\rrbracket$ link. Each node also is equipped with a memory with mean lifetime  $\tau_i$.
	
	\subsection{Performance of Bell State Measurements (BSM)}
	
	The entanglement distribution protocols rely on direct distribution of pairs over short distances and swap those half pairs received at the intermediate nodes in order to realize  a long distance entanglement. These swaps are Bell State Measurements (BSM) of two photons in one node, each of these photons belonging to  different ebit pairs. 
	
	In practice, a simple BSM is limited\footnote{By Hong-Ou-Mandel (HOM) interferences.} and gives  a success probability of $p_{swap} = 0.5$. More complex set-ups have been studied and improve the probability up to $0.6$ \cite{Bayerbach_2023} while theory expects that $p_{swap}$ could reach $0.9375$ \cite{Kilmer_2019,Ewert_2014,Qi2025}. The complexity of these set-ups may raise implementation issues in real networks. As the theoretical framework for \cite{Bayerbach_2023} expects a success probability $p_{swap} = 0.75$ \cite{Ewert_2014}, we will adopt this value in the following of this paper.
	
	Swaps not only are stochastic, but they also lower the fidelity of the final pair as a function of the fidelity of the two input pairs. Consider two pairs with fidelity $F_1$ and $F_2$, the fidelity of the swapped pair is  \cite{Munro15} 
	\begin{equation}
		f_{swap}(F_1, F_2) = F_1F_2 + (1 - F_1)(1-F_2)/3.\label{eq:F_swap}
	\end{equation}
	
	
	\section{Blind Swapping (BS) and Globally Heralded Distribution (GHD)}
	\label{part:BS_GHD}
	
	We first consider two existing protocols, namely Blind Swapping (BS) based on \textit{swap asap} and Globally Heralded Distribution (GHD) which is referred to as Unheralded Swapping in \cite{Abane25}. We consider that time is slotted and denote by $\Delta$ the duration of a slot.
	
	\subsection{Blind Swappping}
	\subsubsection{Principle}
\begin{figure}[htpb]
		\centering
		\includegraphics[width=0.8\linewidth]{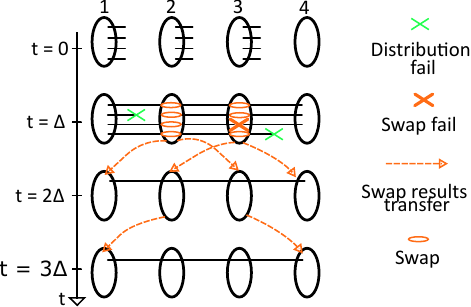}
		\caption{Blind Swapping scheme for four nodes and a width of $n = 4$ generations}
		\label{fig:scheme_BS}
	\end{figure}

	The BS protocol, illustrated in Figure~\ref{fig:scheme_BS}, is totally blind and acts without knowledge of other nodes. We consider a path $r$ composed of $l_r$ links and  memories are used in  nodes. A total number of $n$ pairs are generated in the $n_r = l_r-1$ first nodes (all except the destination). One photon from each pair is immediately sent from node $i \in \llbracket 1, n_r\rrbracket$ to node $i+1$ at a distance $L_i$ through a fiber with attenuation $\alpha_i$. A repeater node expects to receive two photons and swaps them at the end of the time slot.
	
	\subsubsection{Link probability}
	Since swap success, distribution probability, and memory life-time are independent we can look at the probabilities in two different stages. First, there is the distribution from node $i$ to $i+1$ with swapping at time $t = \Delta$ at node $i+1$ except for the destination node $l_r$ which does not perform swapping.
	
	The  probability that the photon is received at node $i+1$ and is still alive at the swapping time writes $$p_i = e^{-\alpha_i L_i}e^{-(\Delta - L_i/c)/\tau_{i+1}}e^{-\Delta/\tau_i}$$ in which we consider the transmission of the photon through the fiber, the time its spends in its destination memory until the end of the time slot, and its pair that stays in its original memory the whole time slot. We can simplify its writing so that it follows
	\begin{equation}
		p_i =e^{-\left(\frac{\Delta}{\overline{\tau}_i}+ \beta_i\right)} \label{eq:p_i}
	\end{equation}
	with  $\frac{1}{\overline{\tau}_i} = \frac{1}{\tau_i} + \frac{1}{\tau_{i+1}}$ and $ \beta_i = L_i(\alpha_i-\frac{1}{c\tau_{i+1}})$. 
	
	Every node needs to properly distribute its photon and they all need to be successfully swapped for a link to be established. As such the probability to establish one link writes \cite{Shi19routing}
	\begin{equation}
		q_{BS} = p_{swap}^{l_r-2}\prod_{i = 1}^{l_r-1}p_i = p_{swap}^{l_r-2}e^{-(\Delta/\overline{\tau} + \beta)}
	\end{equation}
	with $ 1/\overline{\tau} = \sum_{i = 1}^{l_r-1}1/\overline{\tau}_i $  and $\beta = \sum_{i = 1}^{l_r-1}\beta_i$.
	
	An additional time is needed for LOCC operations (Local Operations and Classical Communication) that transmit the results of all swaps to both the source and destination nodes. If we limit one transmission between two first neighbors to last one time slot $\Delta$, then the farthest node would need a time $(l_r-2)\Delta$ to reach one of the two endpoints. Both directions are crossed in parallel times. Since the photons are all in the source and destination nodes now, we can write
	\begin{equation}
		q_{BS} = p_{swap}^{l_r-2}e^{-(\Delta/\overline{\tau} + \beta)}e^{-\Delta (l_r-2)(\frac{1}{\tau_1}+\frac{1}{\tau_{l_r}})}.
	\end{equation}
	
	The probability to establish a link follows a Bernoulli law of parameter $q_{BS}$. If we reiterate the process $n$ times (referred to as the width of the path), the number of end-to-end entanglements follows is a binomial random variable $\mathcal{B}(n,q_{BS})$, with mean $nq_{BS}$ and variance $nq_{BS}(1-q_{BS})$
	
	\subsubsection{Link fidelity}
	
	Throughout the protocol, the pairs fidelity decrease. In order to write the link fidelity, we use Equations \eqref{eq:F_decoh} and \eqref{eq:F_swap}. It proceeds as follows: 
	\begin{enumerate}
		\item The pairs are generated with fidelity 1.
		\item Each pair from node $i\in\llbracket 1,n_r\rrbracket$ is then distributed for a time $L_i/c$ and then stored during $\Delta - L_i/c$ so that the initial fidelity writes
		\begin{equation}
			F_{0i} = f_{\tau}(\Delta-\frac{L_i}{c},\min(f_{\tau_{coh}}(\frac{L_i}{c},1),f_{\tau}(\frac{L_i}{c},1))).
		\end{equation}
		\item The index $0i$ denotes the initial state from node $i$. For each node we may define a swap function $s_i : F \mapsto f_{swap}(F,F_{0i})$, so that after swapping at all nodes at the same time the fidelity of the resulting links is the iterated quantity $F = \underset{i=2}{\overset{n_r}{\circ}}s_i(F_{01})$.
		\item The protocol finishes after waiting for a time equal to $(l_r-2)\Delta$ so that at the end, the link fidelity writes
		\begin{equation}
			F = f_{\tau}((l_r-2)\Delta, \underset{i=2}{\overset{n_r}{\circ}}s_i(F_{01})).\label{eq:F_final}
		\end{equation}
	\end{enumerate}
	
	This protocol is the most basic one. In the next subsection, we study a more elaborated protocol that should be more efficient.
	
	\subsection{Globally Heralded Distribution (GHD)}
	
	The Globally Heralded Distribution protocol (GHD) is an extension of the Unheralded Swapping protocol from \cite{Abane25}. The Unheralded Swapping uses heralded distribution, in which all the swapping nodes only perform the minimal amount of swapping. Each node swaps without any knowledge of the other nodes swapping results, which is why it was originally called Unheralded Swapping.
	
	We use the Unheralded Swapping concept and add the memories probabilities to it; in this case, all the nodes have to know the minimum number of swaps to perform.
	
	\subsubsection{Principle}
	
\begin{figure}[htpb]
		\centering
		\includegraphics{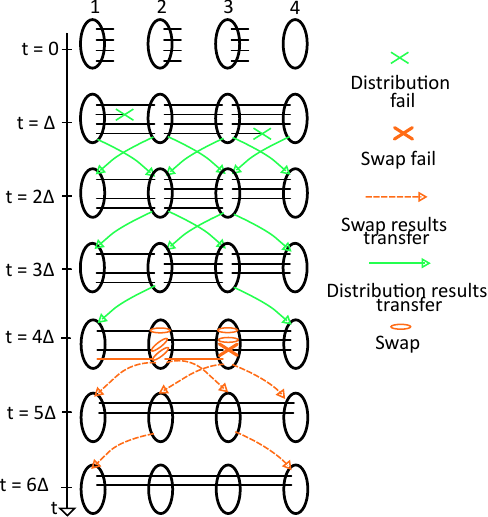}
		\caption{Globally Heralded Swapping  scheme for four nodes and a width of $n = 4$ generations}
		\label{fig:scheme_GHD}
	\end{figure}

	The GHD described in Figure~\ref{fig:scheme_GHD} makes smarter swapping decisions at the expense of longer time periods spent in quantum memories. It will indeed only swap pairs it knows have fully been distributed, and only the minimal amount that has been distributed over all nodes. For instance, on a path of length $l_r = 3$, if the first node has distributed $3$ pairs and the second one $2$, then only two swaps will be done at the second node.
	
	For the data of all nodes to reach one deciding node and then go back to all nodes, it takes \textit{a minima} $l_r-1$ time slots.
    After swapping, an additional time $(l_r-2)\Delta$ is needed for LOCC, in the same fashion as for BS.
	
	\subsubsection{Link probability}
	We write the entanglement success as a decomposition of success events knowing the minimum number of pairs available for swapping. If we denote by $M_{l_r}^n$ that minimum, the probability to establish exactly $k\in\llbracket 0,n\rrbracket$ links is
	\begin{equation}
		\P(e2e = k) = \sum_{m = k}^{n}\P(e2e_{GHD} = k | M_{l_r}^n = m)\P(M_{l_r}^n = m).
	\end{equation}
	
	That minimum probability is expressed with a recurrence formula in \cite{Abane25}. We have derived a direct formula, allowing lower time complexity: since
	$$
	\P(M_{l_r}^n > k) = \prod_{i = 1}^{n_r}\P(S_i^n > k),
	$$
	we have
	\begin{equation}
		\P(M_{l_r}^n = m) = \P(M_{l_r}^n > m-1) - \P(M_{l_r}^n > m).
		\label{eq:P_min}
	\end{equation}
	
	
	We denote for $ i\in\llbracket 1, l_r-1\rrbracket$, $S_i$ the event that counts the number of successful distributions from node $i$ to node $i+1$. For a width of $n$ generations, it follows a binomial law $\mathcal{B}(n,p_i)$ with $p_i$ from \eqref{eq:p_i}. The probability to establish $k\in\llbracket 0,n\rrbracket$ links, knowing the minimum amount of swaps $m\in\llbracket 0,n\rrbracket$, is a binomial random variable  $\mathcal{B}(m,q_{GHD})$.
	
	The elementary success of the link establishment attempt requires that all swaps succeed, with a probability $p_{swap}^{l_r-2}$. We also need that the pairs at all nodes survive the waiting time $(l_r-1)\Delta$, which is independent from the swap success. Hence, we have
    	\begin{equation}
		q_{GHD} = p_{swap}^{l_r-2}e^{-\Delta(\frac{1}{\overline{\tau}}(l_r-1) + (l_r-2)(\frac{1}{\tau_1}+\frac{1}{\tau_{l_r}}))},
	\end{equation} 
	where the last part of the exponential represents the LOCC delay at the end of the protocol. 
	
	\subsubsection{Link fidelity}
	The final link fidelity follows the same Equation~\eqref{eq:F_final} as for BS, with a difference in the initial fidelity. It needs the additional waiting time $(l_r-1)\Delta$ so that for all $i\in\llbracket 1,l_r-1\rrbracket$, 
    	$$
		F_{0i}= f_{\tau}\left(\Delta l_r-\frac{L_i}{c},\right. \\ \left.
		\min\left(f_{\tau_{coh}}\left(\frac{L_i}{c},1\right),f_{\tau}\left(\frac{L_i}{c},1\right)\right)\right).
	$$
	
	The GHD protocol has  communication times that scale with the length of the route. This in turns impacts the link fidelity and success probability of entanglement distribution; this  is why we have elaborated a modified version with a fixed communication time.
	
	\section{Locally Heralded Distribution (LHD)}
	\label{part:LHD}
	\subsection{Principle}
	
\begin{figure}[htpb]
		\centering
		\includegraphics{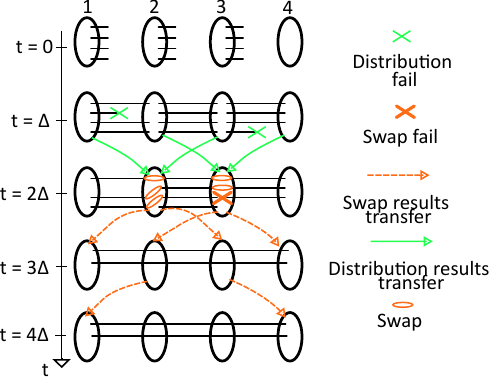}
		\caption{Locally Heralded Swapping  scheme for four nodes and a width of $n = 4$ generations}
		\label{fig:scheme_LHD}
	\end{figure}

	From the previous section, we observe that  swapping with knowledge reduces the amount of communication and the consumption of entangled pairs. Hence, in the LHD protocol illustrated in Figure~\ref{fig:scheme_LHD}, the swap operations are performed with information from the nearest neighbors only. As a consequence, the number of swaps to be operated at each node is a local minimum determined by the nearest neighbors. 
	
	The information received from the neighbors concerns distribution results, namely which photons have been received and which of the stored photons have been lost.  The duration of this data transfer is fixed and only lasts one time slot $\Delta$, as opposed to the duration for GHD which is proportional to the length of the route.
	
	\subsection{Link probability}
	
	The link probability of LHD follows the same law as that of GHD, with the difference of the time spent in memory. Thus $\forall k\in\llbracket 0,n\rrbracket$,
	\begin{equation}
		\P(e2e = k) = \sum_{m = k}^{n}\P(e2e_{LHD} = k | M_{l_r}^n = m)\P(M_{l_r}^n = m).
	\end{equation} with $\forall m\in\llbracket 0,n\rrbracket$ the event $e2e_{LHD}|M_{l_r}^n = m$ following the law $\mathcal{B}(m,q_{LHD})$ and \begin{equation}
		q_{LHD} = p_{swap}^{l_r-2}e^{-\Delta(\frac{1}{\overline{\tau}} + (l_r-2)(\frac{1}{\tau_1}+\frac{1}{\tau_{l_r}}))}.
	\end{equation}
	\subsection{Link fidelity}
	
	Following the same process, the link fidelity of LHD follows \eqref{eq:F_final} with $\forall i\in\llbracket 1,n_r\rrbracket$, \begin{equation}
		F_{0i} = f_{\tau}(2\Delta-\frac{L_i}{c},
		\min(f_{\tau_{coh}}(\frac{L_i}{c},1),f_{\tau}(\frac{L_i}{c},1))).
	\end{equation}
	
	\section{Performance analysis}
	\label{performance}
\begin{figure}[hbtp]
						 \scalebox{.8}{\includegraphics{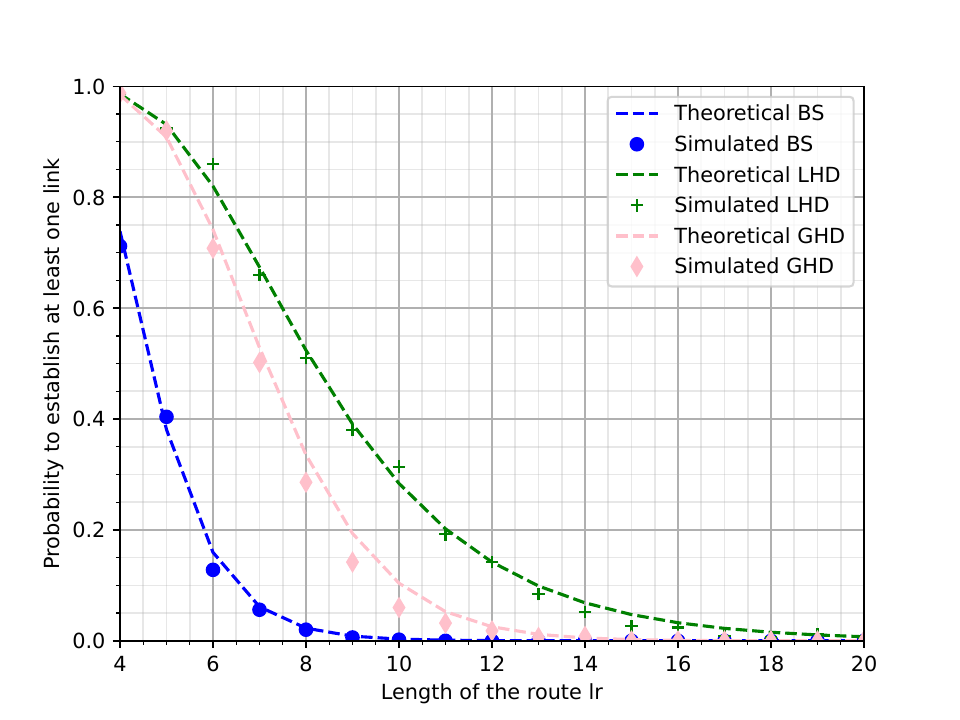}}           
		\caption{Link probability for the three protocols as a function of the route length for a width of $n = 20$ generations and a distance between nodes of $L = 50km$. The attenuation parameter is $\alpha = 0.14dB/km$ with a celerity through fiber of $c = 220.10^{6}m.s^{-1}$.}
		\label{fig:P}
	\end{figure}

	We plot the link probabilities for all three protocols as a function of the route length. As a matter of comparison we coded a simulation on python for each protocol. The memory life time is chosen to be $\tau = 200ms$, identical for all nodes, which is on the higher end of the contemporary possibilities. From Figure~\ref{fig:P} we observe that the simulation is in accordance with theory.
	
	There is a clear advantage of LHD over the two others, especially when $l_r$ increases. The global protocol (GHD) shows great advantage over the naseline BS, but its efficiency sinks to zero almost twice faster than its local counter part: $P(e2e_{GHD} \geq 1)\approx 0$ for $l_r = 12$ whereas $P(e2e_{LHD}\geq 1) > 0$ for $l_r = 18$. The difference between LHD and GHD is explained by the different memory storage durations.
	
	\begin{figure}[hbtp]
	
		\scalebox{.8}{\includegraphics{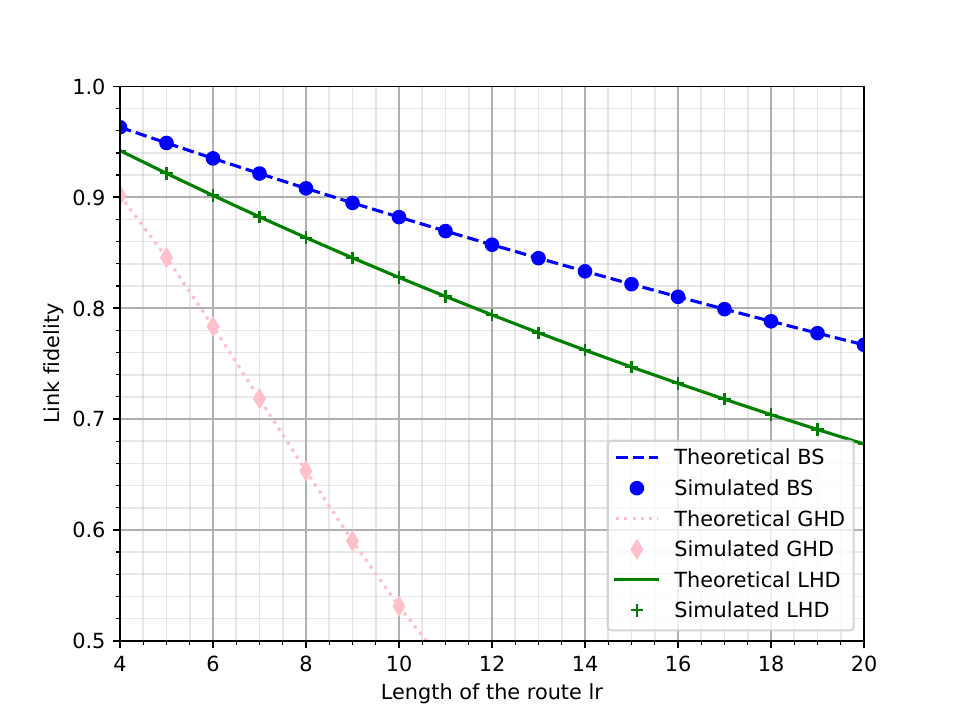}}
		\caption{Link fidelity as a function of the route length $l_r$ for a time slot $\Delta = 2ms$.}
		\label{fig:F}
	\end{figure}
	
	We can also observe in Figure~\ref{fig:F} that GHD has a dramatic decrease in fidelity when the number of nodes on the path increases. Our protocol LHD offers a great improvement in this respect. The link fidelity of LHD follows a  trend similar to that of BS, the latter being overall better. Nevertheless, if we compare link probabilities, LHD is much better than BS. This means that in order to reach the same link probability, BS requires a much greater number of pairs of entangled pairs than LHD. It could then be possible to use more pairs for distillation protocols in LHD and still be more efficient than BS. Our protocol could hence reach both high link fidelity and probability.
	
	\section{Conclusion}
	\label{conclusion}
	\balance
	Inspired by the GHD protocol, we have introduced in the paper the so-called heralded distribution (LHD) protocol. This latter yields a success probability as good as its global counterparts for perfect memories and outperforms it for imperfect ones. 
	
	The longer protocol duration of LHD compared with Blind Swapping makes the final link fidelity lower, but the high number of links that may be created at a lower cost could permit purification protocols. Additionally, the LHD could be improved for a higher link fidelity. 
	A faster protocol, with better success probability and less classical overhead, would achieve a better resource management. This will be addressed in further studies.

    \section*{Acknowledgments}

We acknowledge the support of the European Union’s Horizon Europe research and innovation program
through the Quantum Secure Network Partnership project (QSNP).
	
	\bibliographystyle{IEEEtran}
	\bibliography{ref}

@article{Dupuy23,
	author = {Dupuy, Fabrice and Goursaud, Claire and Guillemin, Fabrice},
	year = {2023},
	month = {03},
	pages = {2200180},
	title = {A Survey of Quantum Entanglement Routing Protocols—Challenges for Wide‐Area Networks},
	volume = {6},
	journal = {Advanced Quantum Technologies},
	doi = {10.1002/qute.202200180}
}

@article{li22fidel,
	title={Fidelity-Guarantee Entanglement Routing in Quantum Networks}, 
	author={Li, Jian and Wang, Mingjun and Xue, Kaiping and Li, Ruidong and Yu, Nenghai and Sun, Qibin and Lu, Jun},
	journal={IEEE Transactions     on Communications}, 
	year={2022},
	volume={70},
	number={10},
	pages={6748-6763},
	doi={10.1109/TCOMM.2022.3200115}
}

@misc{Shi19routing,
	title={Modeling and Designing Routing Protocols in Quantum Networks}, 
	author={Shouqian Shi and Chen Qian},
	year={2019},
	eprint={1909.09329},
	archivePrefix={arXiv},
	primaryClass={cs.NI},
	url={https://arxiv.org/abs/1909.09329}, 
}

@article{Munro15,
	author={Munro, William J. and Azuma, Koji and Tamaki, Kiyoshi and Nemoto, Kae},
	journal={IEEE Journal of Selected Topics in Quantum Electronics}, 
	title={Inside Quantum Repeaters}, 
	year={2015},
	volume={21},
	number={3},
	pages={78-90},
	doi={10.1109/JSTQE.2015.2392076}}

@article{Wehner18,
	author = {Stephanie Wehner  and David Elkouss  and Ronald Hanson },
	title = {Quantum internet: A vision for the road ahead},
	journal = {Science},
	volume = {362},
	number = {6412},
	pages = {eaam9288},
	year = {2018},
	doi = {10.1126/science.aam9288},
	URL = {https://www.science.org/doi/abs/10.1126/science.aam9288},
	eprint = {https://www.science.org/doi/pdf/10.1126/science.aam9288}}

@article{Beauchamp25,
	title={A Modular Quantum Network Architecture for Integrating Network  Scheduling with Local Program Execution},
	author={Thomas R. Beauchamp and Hana Jirovská and Scarlett Gauthier},
	year={2025},
	month={3},
	journal = {arXiv},
	eprint={2503.12582},
	archivePrefix={arXiv},
	primaryClass={cs.NI}
}

@article{Kilmer_2019,
	title={Boosting linear-optical Bell measurement success probability with predetection squeezing and imperfect photon-number-resolving detectors},
	volume={99},
	ISSN={2469-9934},
	url={http://dx.doi.org/10.1103/PhysRevA.99.032302},
	DOI={10.1103/physreva.99.032302},
	number={3},
	journal={Physical Review A},
	publisher={American Physical Society (APS)},
	author={Kilmer, Thomas and Guha, Saikat},
	year={2019},
	month={3} }

@article{Abane25,
	author = {Abane, Amar and Cubeddu, Michael and Mai, Van and Battou, Abdella},
	year = {2025},
	month = {01},
	pages = {1-36},
	title = {Entanglement Routing in Quantum Networks: A Comprehensive Survey},
	volume = {PP},
	journal = {IEEE Transactions on Quantum Engineering},
	doi = {10.1109/TQE.2025.3541123}
}

@article{Komar_2014,
	title={A quantum network of clocks},
	volume={10},
	ISSN={1745-2481},
	url={http://dx.doi.org/10.1038/nphys3000},
	DOI={10.1038/nphys3000},
	number={8},
	journal={Nature Physics},
	publisher={Springer Science and Business Media LLC},
	author={Kómár, P. and Kessler, E. M. and Bishof, M. and Jiang, L. and Sørensen, A. S. and Ye, J. and Lukin, M. D.},
	year={2014},
	month={6}, pages={582–587} }

@article{Ewert_2014,
	title={3/4-Efficient Bell Measurement with Passive Linear Optics and Unentangled Ancillae},
	volume={113},
	ISSN={1079-7114},
	url={http://dx.doi.org/10.1103/PhysRevLett.113.140403},
	DOI={10.1103/physrevlett.113.140403},
	number={14},
	journal={Physical Review Letters},
	publisher={American Physical Society (APS)},
	author={Ewert, Fabian and van Loock, Peter},
	year={2014},
	month={9}}

@article{Bayerbach_2023,
	author = {Matthias J. Bayerbach  and Simone E. D’Aurelio  and Peter van Loock  and Stefanie Barz },
	title = {Bell-state measurement exceeding 50\% success probability with linear optics},
	journal = {Science Advances},
	volume = {9},
	number = {32},
	pages = {eadf4080},
	year = {2023},
	doi = {10.1126/sciadv.adf4080},
	URL = {https://www.science.org/doi/abs/10.1126/sciadv.adf4080},
	eprint = {https://www.science.org/doi/pdf/10.1126/sciadv.adf4080}}

@article{Qi2025,
	author = {Qi, Ji and Yu, Chang-Qi and Yuan, Rui-Yang and Yang, Zhe and Ren, Bao-Cang},
	title = {Entanglement-assisted logical Bell state measurement with linear optics},
	doi = {10.1016/j.optlastec.2025.113385},
	journal = {Opt. Laser Tech.},
	volume = {192},
	pages = {113385},
	year = {2025}}

@misc{Pirker2025,
	title={A resource-centric, task-based approach to quantum network control}, 
	author={Alexander Pirker and Belen Munoz and Wolfgang Dür},
	year={2025},
	eprint={2507.12030},
	archivePrefix={arXiv},
	primaryClass={quant-ph},
	url={https://arxiv.org/abs/2507.12030}, 
}

@phdthesis{Coopmans2024,
	author = {T. J. Coopmans},
	title = {Tools for the design of quantum repeater networks},
	type = {Doctoral thesis},
	school = {TU Delft - QID/Elkouss Group},
	year = {2021},
	OPTpagetotal = {242},
	OPTdoi = {https://doi.org/10.4233/uuid:90d06f1d-4f23-48cc-8f96-51500258020f}
}

@misc{Ahmed:2025,
	author = "Ahmed, Shakil and Saeed, Muhammad Kamran and Khokhar, Ashfaq",
	title = "{OSI Stack Redesign for Quantum Networks: Requirements, Technologies, Challenges, and Future Directions}",
	eprint = "2506.12195",
	archivePrefix = "arXiv",
	primaryClass = "quant-ph",
	month = "6",
	year = "2025"
}

@misc{chakraborty2019,
	title={Distributed Routing in a Quantum Internet}, 
	author={Kaushik Chakraborty and Filip Rozpedek and Axel Dahlberg and Stephanie Wehner},
	year={2019},
	eprint={1907.11630},
	archivePrefix={arXiv},
	primaryClass={quant-ph},
	url={https://arxiv.org/abs/1907.11630}, 
}

@article{Gyongyosi_2019,
	title={Adaptive routing for quantum memory failures in the quantum Internet},
	volume={18},
	ISSN={1573-1332},
	url={http://dx.doi.org/10.1007/s11128-018-2153-x},
	DOI={10.1007/s11128-018-2153-x},
	number={2},
	journal={Quantum Information Processing},
	publisher={Springer Science and Business Media LLC},
	author={Gyongyosi, Laszlo and Imre, Sandor},
	year={2019},
	month=jan }

@article{Gyongyosi_2018,
	title={Decentralized base-graph routing for the quantum internet},
    author={Gyongyosi, Laszlo and Imre, Sandor},
	volume={98},
	ISSN={2469-9934},
	url={http://dx.doi.org/10.1103/PhysRevA.98.022310},
	DOI={10.1103/physreva.98.022310},
	number={2},
	journal={Physical Review A},
	publisher={American Physical Society (APS)},
	year={2018},
	month=aug }

@article{Gyongyosi_2017,
	title={Entanglement-Gradient Routing for Quantum Networks},
	volume={7},
	ISSN={2045-2322},
	url={http://dx.doi.org/10.1038/s41598-017-14394-w},
	DOI={10.1038/s41598-017-14394-w},
	number={1},
	journal={Scientific Reports},
	publisher={Springer Science and Business Media LLC},
	author={Gyongyosi, Laszlo and Imre, Sandor},
	year={2017},
	month=oct }

@inproceedings{ZhaoY2022,
	author = {Zhao, Yangming and Zhao, Gongming and Qiao, Chunming},
	title = {E2E Fidelity Aware Routing and Purification for Throughput Maximization in Quantum Networks},
	year = {2022},
	publisher = {IEEE Press},
	url = {https://doi.org/10.1109/INFOCOM48880.2022.9796814},
	doi = {10.1109/INFOCOM48880.2022.9796814},
	booktitle = {IEEE INFOCOM 2022 - IEEE Conference on Computer Communications},
	pages = {480–489},
	numpages = {10},
	location = {London, United Kingdom}
}

@article{Chakraborty_2020,
	title={Entanglement Distribution in a Quantum Network: A Multicommodity Flow-Based Approach},
	volume={1},
	ISSN={2689-1808},
	url={http://dx.doi.org/10.1109/TQE.2020.3028172},
	DOI={10.1109/tqe.2020.3028172},
	journal={IEEE Transactions on Quantum Engineering},
	publisher={Institute of Electrical and Electronics Engineers (IEEE)},
	author={Chakraborty, Kaushik and Elkouss, David and Rijsman, Bruno and Wehner, Stephanie},
	year={2020},
	pages={1–21} }

@misc{Patil2024,
	title={Entanglement Routing using Quantum Error Correction for Distillation}, 
	author={Ashlesha Patil and Michele Pacenti and Bane Vasić and Saikat Guha and Narayanan Rengaswamy},
	year={2024},
	eprint={2405.00849},
	archivePrefix={arXiv},
	primaryClass={quant-ph},
	url={https://arxiv.org/abs/2405.00849}, 
}

@article{Duer99,
	title = {Quantum repeaters based on entanglement purification},
	author = {Duer, W. and Briegel, H.-J. and Cirac, J. I. and Zoller, P.},
	journal = {Phys. Rev. A},
	volume = {59},
	issue = {1},
	pages = {169--181},
	numpages = {0},
	year = {1999},
	month = {1},
	publisher = {American Physical Society},
	doi = {10.1103/PhysRevA.59.169},
	url = {https://link.aps.org/doi/10.1103/PhysRevA.59.169}
}

@INPROCEEDINGS{Wang24_SES_noF_nor_mem,
  author={Wang, Zhaoying and Li, Jian and Li, Zhonghui and Chen, Lutong and Yu, Nenghai and Sun, Qibin and Lu, Jun},
  booktitle={2024 International Conference on Quantum Communications, Networking, and Computing (QCNC)}, 
  title={Efficient Routing Design Based on Entanglement Flow Loss Effect in Quantum Networks}, 
  year={2024},
  volume={},
  number={},
  pages={9-16},
  doi={10.1109/QCNC62729.2024.00032}}

@misc{Hughes:2025_asynchro_SES_<_PES,
    author = {Hughes, Emma and Munizzi, William and Narang, Prineha},
    title = {A Compact Framework for Analyzing Asynchronous Entanglement Distribution in Quantum Networks},
    eprint = {2507.22992},
    archivePrefix = {arXiv},
    primaryClass = {quant-ph},
    month = {7},
    year = {2025}
}

@inproceedings{Zang_2023_F_decoh_mem,
   title={Entanglement Distribution in Quantum Repeater with Purification and Optimized Buffer Time},
   url={http://dx.doi.org/10.1109/INFOCOMWKSHPS57453.2023.10226122},
   DOI={10.1109/infocomwkshps57453.2023.10226122},
   booktitle={IEEE INFOCOM 2023 - IEEE Conference on Computer Communications Workshops (INFOCOM WKSHPS)},
   publisher={IEEE},
   author={Zang, Allen and Chen, Xinan and Kolar, Alexander and Chung, Joaquin and Suchara, Martin and Zhong, Tian and Kettimuthu, Rajkumar},
   year={2023},
   month=may, pages={1–6} }

@article{Inesta_2023,
	title={Optimal entanglement distribution policies in homogeneous repeater chains with cutoffs},
	volume={9},
	ISSN={2056-6387},
	url={http://dx.doi.org/10.1038/s41534-023-00713-9},
	DOI={10.1038/s41534-023-00713-9},
	number={1},
	journal={npj Quantum Information},
	publisher={Springer Science and Business Media LLC},
	author={I\~nesta, \'Alvaro G. and Vardoyan, Gayane and Scavuzzo, Lara and Wehner, Stephanie},
	year={2023},
	month={5} }

@article{Kumar_Pankaj_2025,
title = {ZBR: Zone-based routing in quantum networks with efficient entanglement distribution},
journal = {Journal of Network and Computer Applications},
volume = {238},
pages = {104156},
year = {2025},
issn = {1084-8045},
doi = {https://doi.org/10.1016/j.jnca.2025.104156},
url = {https://www.sciencedirect.com/science/article/pii/S1084804525000530},
author = {Pankaj Kumar and Binayak Kar}
}

@misc{Kumar_Vinay_2025,
      title={Routing in Quantum Networks with End-to-End Knowledge}, 
      author={Vinay Kumar and Claudio Cicconetti and Marco Conti and Andrea Passarella},
      year={2025},
      eprint={2407.14407},
      archivePrefix={arXiv},
      primaryClass={quant-ph},
      url={https://arxiv.org/abs/2407.14407}, 
}

@misc{Kar2023,
      title={Routing Protocols for Quantum Networks: Overview and Challenges}, 
      author={Binayak Kar and Pankaj Kumar},
      year={2023},
      eprint={2305.00708},
      archivePrefix={arXiv},
      primaryClass={quant-ph},
      url={https://arxiv.org/abs/2305.00708}, 
}

@article{Shi2024_QCAST,
author = {Shi, Shouqian and Zhang, Xiaoxue and Qian, Chen},
year = {2024},
month = {06},
pages = {1-16},
title = {Concurrent Entanglement Routing for Quantum Networks: Model and Designs},
volume = {PP},
journal = {IEEE/ACM Transactions on Networking},
doi = {10.1109/TNET.2023.3343748}
}

@article{Pirandola_2017,
   title={Fundamental limits of repeaterless quantum communications},
   volume={8},
   ISSN={2041-1723},
   url={http://dx.doi.org/10.1038/ncomms15043},
   DOI={10.1038/ncomms15043},
   number={1},
   journal={Nature Communications},
   publisher={Springer Science and Business Media LLC},
   author={Pirandola, Stefano and Laurenza, Riccardo and Ottaviani, Carlo and Banchi, Leonardo},
   year={2017},
   month={4}}

@article{Welinq_QM,
author = {Hadriel Mamann  and Thomas Nieddu  and Félix Hoffet  and Mathieu Bozzio  and Félix Garreau de Loubresse  and Iordanis Kerenidis  and Eleni Diamanti  and Alban Urvoy  and Julien Laurat },
title = {Quantum cryptography integrating an optical quantum memory},
journal = {Science Advances},
volume = {11},
number = {38},
pages = {eadx3223},
year = {2025},
doi = {10.1126/sciadv.adx3223},
URL = {https://www.science.org/doi/abs/10.1126/sciadv.adx3223},
eprint = {https://www.science.org/doi/pdf/10.1126/sciadv.adx3223}}
    
\end{document}